\documentstyle[sprocl]{article}

\input{psfig.sty}

\def\Journal#1#2#3#4{{#1} {\bf #2}, #3 (#4)}

\begin{document}

\title{HALO TRACING WITH ATOMIC HYDROGEN}

\author{MICHAEL R. MERRIFIELD}

\address{School of Physics \& Astronomy, University of Nottingham\\
Nottingham NG7 2RD\\
UK\\
E-mail: michael.merrifield@nottingham.ac.uk}


\maketitle\abstracts{This paper reviews the constraints that can be
placed on the shapes of disk galaxies' dark halos using the
distribution and kinematics of atomic hydrogen.  These data indicate
that dark halos are close to axisymmetric, with their axes of symmetry
co-aligned with their disk axes.  They also appear to be oblate, with
shortest-to-longest axis ratios displaying quite a broad range of values
from $\sim 0.2$ to $\sim 0.8$.  These results are consistent with the
predicted shapes of halos in cold dark matter scenarios, but rule out
some of the more exotic dark matter candidates.  However, the total
number of measurements is still depressingly small, and more data are
required if halo shape is to become a powerful diagnostic for theories
of galaxy formation and evolution.}

\section{Introduction}
Atomic hydrogen (HI) is ubiquitous in the Universe, and the convenient
21cm line means that its kinematics are readily determined, making it
an ideal tracer for dynamical studies of the mass distributions of
galaxies.  The simplest application lies in determining the radial
mass profiles in galaxies using their HI rotation curves -- indeed, it
was through such studies that the existence of dark matter in these
systems was first credibly demonstrated.\cite{k87} Taking such
analyses a stage further, one can start to study the shape of the mass
distribution, and hence infer something about the shape of the dark
halo.  Clearly, this further step requires one to relax the assumption
of spherical symmetry in the analysis, and such a step always
complicates the analysis.  Further, the hydrodynamical nature of the
gas can be conveniently ignored when one treats it as a thin
axisymmetric disk of material on circular orbits.  If one relaxes the
imposed symmetry, the orbits of the material in what may now be a
triaxial potential become much more complex, making it likely that
hydrodynamically-dictated collisions between different components will
occur.  In addition, since one is now interested in the full
three-dimensional structure of the galaxy, one cannot just consider
the distribution in the direction in which the gas disk is primarily
centrifugally supported; the hydrodynamical processes that support the
gas distribution in other directions must also be included in the
analysis.

Despite these complexities, we can still obtain at least crude
measures of the full three-dimensional shapes of disk galaxies' halos
from HI studies.  In Section~\ref{sec:par}, we look at the constraints
that such analyses place on the shapes of halos in the planes of disk
galaxies, while Section~\ref{sec:perp} looks at the corresponding data
perpendicular to their disk planes.

\section{Halo Shapes in the Planes of Galaxies}\label{sec:par}
\begin{figure}[t]
\hfill \psfig{figure=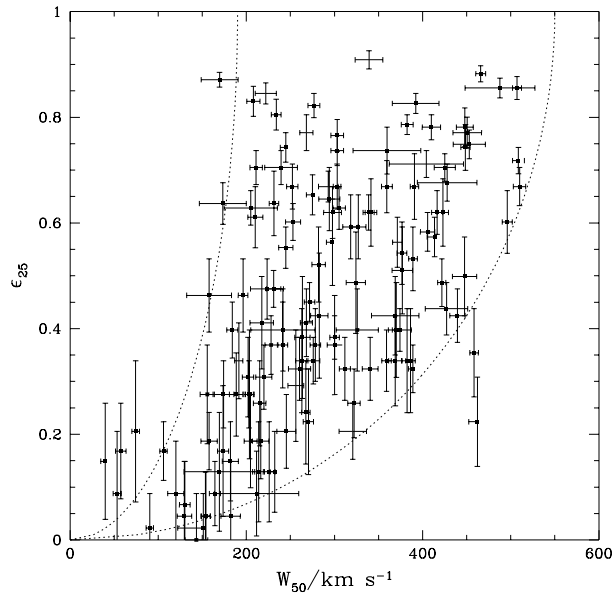,height=3.3in} \hfill{} 
\caption{The ellipticities of disk galaxies as determined at the $25\,{\rm
mag}\,{\rm arcsec}^{-2}$ isophote, plotted against the full width half
maximum of their HI line widths.  The data were drawn from the LEDA
database, and consist of all galaxies of types Sa -- Sc that are
bright in HI ($M_{21} < 13$), and that have their axis ratios
determined to better than 10\%.  Since ellipticity is defined to be
positive, there will be a bias in this plot that increases the mean
ellipticity where it is close to zero.  The lines show where thin
axisymmetric disks with two different rotation speeds should lie on
this plot. \label{fig:hiwidths}}
\end{figure}

As mentioned above, the key to studying galaxies' mass distributions
is that the orbits of material are dictated by the gravitational
potential.  If the mass distribution is not spherically symmetric, one
would expect this to show up as a corresponding absence of symmetry in
the orbits.  One problem in studying disk galaxies is that it is not
entirely straightforward to establish their symmetry: a disk that
appears elliptical may do so because it is intrinsically
non-symmetric, but it could equally be an axisymmetric disk that is
not viewed face on.  Thus, the simplest use of HI in this context is
to establish a disk's orientation, by using the fact that the width of
the 21cm emission integrated from the entire galaxy depends on how
close the galaxy is to face on.  Figure~\ref{fig:hiwidths}\ shows a
simple implementation of this approach, where I have plotted apparent
optical ellipticity versus HI line width for a sample of nearby
galaxies.  Those galaxies with very small line widths, which must lie
close to face-on, all have very low ellipticities, implying that disk
galaxies are close to round in shape, and hence that the potential in
the disk plane is nearly axisymmetric.  Fitting this data distribution
with an elliptical disk model places a constraint on the mean
ellipticity in the gravitational potential of $\epsilon < 0.1$.

In fact, localized star formation, spiral arms, etc, may render a
face-on globally-circular disk somewhat elliptical in appearance, so
this plot can really only provide an upper limit on the true
ellipticities of disks; a better analysis\cite{rz95} using near
infrared data to avoid most of this problem places an even tighter
constraint on the ellipticity of the potential in the plane, yielding
a value of $0.045^{+0.03}_{-0.02}$.

A related constraint can be derived by noting the tightness of the
Tully-Fisher relation.  Using infrared photometry, the relationship
between the inclination-corrected HI line widths and absolute
magnitudes of disk galaxies shows a scatter of only $\sim 0.25$
magnitudes.\cite{getal97} If disk galaxies were intrinsically
elliptical, then there would be two new contributions to the scatter:
first, if the apparent ellipticity were used to estimate a galaxy's
inclination, then non-circular disks would lead to erroneous
inclination corrections in the HI line width; and second, the
non-circular motions of the HI gas would mean that the relationship
between observed line width and average circular speed would vary
depending on the orientation of the disk's major axis, again blurring
the relation.  A study of these effects\cite{fdz92} placed a limit on
the average ellipticity of the potential in the planes of disk
galaxies of $\epsilon < 0.1$.  Such an analysis complements what we
learn from the optical disk shape analysis, since the HI disks
generally lie at larger radii: we thus have some indication that the
shape of the mass distribution does not vary significantly over a
range in radii.

In addition to these global techniques for modeling the average
properties of the gravitational potentials of an ensemble of galaxies,
one can also study the kinematics of HI in individual galaxies in
order to measure their halo shapes.  In particular, the HI rings found
around some galaxies make exceptional probes of these systems' shapes.
For example, a study\cite{fvgdz94} of the HI gas ring around the
early-type galaxy IC~2006 showed that the line-of-sight velocity as a
function of azimuth around the ring could be accurately modeled if the
ring were intrinsically circular.  Formally, this analysis measures
the ellipticity of the potential at the radius of the ring as
$\epsilon = 0.012 \pm 0.026$.  These techniques can be extended to
studies of complete disks of HI or H$\alpha$ emission, essentially by
fitting the gas intensity and line-of-sight velocity at each point
using a series of tilted rings.  Such studies\cite{sfdz97,aetal01}
also indicate that the potential is close to axisymmetric in the disk
plane of normal spiral galaxies, with typical values of $\epsilon \sim
0.05$, although there seem to be a few galaxies with ellipticities as
large as $\epsilon \sim 0.2$.\cite{aetal01}

\section{Halo Shapes Perpendicular to the Planes of Galaxies}\label{sec:perp}
Perpendicular to the planes of disk galaxies, constraints on the
shapes of halos are rather harder to come by.  One obvious approach is
to look at polar ring galaxies, in which a ring of material, usually
containing HI, orbits around the pole of a disk galaxy.  In such a
fortuitous arrangement, one can use the type of analysis described
above to model the kinematics of the ring, and hence assess the shape
of the potential in this direction.  Although such analyses provide
some of the most reliable information on the shapes of
halos,\cite{s01} there is one important caveat associated with these
studies: polar ring systems are not normal galaxies.  The ring
probably formed in a significant merger, and it is quite likely that
such a cataclysmic event will also alter the structure of the halo in
the resulting composite system.  It is therefore not obvious that the
shapes of halos in these systems are representative of the general
population. 

A more generally applicable technique involves looking at the
distribution of gas in a disk galaxy perpendicular to its
plane.\cite{o95} The thickness of the gas layer is dictated by the
hydrostatic balance between the internal turbulent motions of the gas,
and the pull of gravity towards the plane: the more mass there is
close to the plane, the thinner the distribution into which the gas is
gravitationally squeezed.  Thus, if one uses the rotation curve to
derive the radial distribution of mass, one can then determine how
closely this mass is concentrated toward the galactic plane by the
observed thickness of the gas layer.  Since the density of the dark
halo drops with radius, the force confining the gas layer decreases,
and so one characteristically sees a ``flaring'' in the HI layer at
large radii; it is exactly how the layer thickness increases with
radius that provides the constraint on halo flattening.

In the mid-1990s, this technique was successfully applied to a few
galaxies,\cite{o95,o96} but there were a number of initial doubts over
its validity.  In particular, the hydrodynamical behavior of the gas
means that the physics is nowhere near as clear cut as for a pure
gravitational problem: other forces such as cosmic ray and magnetic
field pressures could well be significant, and it is not even obvious
quite what velocity dispersion should be associated with the turbulent
motions of gas clouds in the hydrostatic equilibrium equation.  These
concerns were compounded when both of the initial applications (to
NGC~891 and NGC~4244) implied highly flattened dark halos with
shortest-to-longest ratios in the density distribution of $\sim 0.3$.
These results were only marginally consistent with those derived using
other observational techniques (such as the study of polar ring
systems), and were not at all what most theories suggested.

In order to test the validity of the method, we decided to apply it to
the Milky Way's HI layer, as in our own galaxy there are other
techniques that one can use to determine the halo shape.  The first
requirement for such an analysis is a reliable measure of the Galaxy's
rotation curve.  In the inner Galaxy, this is readily determined from
HI data by the tangent point method, but at Galactic radii larger than
the Sun's (where the crucial flaring in the gas layer occurs), this
method is not available.  Fortunately, an alternative technique has
been developed,\cite{m92} which assumes only that the thickness of the
Galactic gas layer does not vary with azimuth around the Galaxy, and
that the material is moving around circular orbits.  With these
assumptions, one can solve simultaneously for both the rotation curve
and the thickness of the layer as a function of radius -- both the
ingredients required for determining the shape of the dark halo.
Applying the gas layer technique to these data resulted in a
measurement of the Milky Way's halo shortest-to-longest ratio of $\sim
0.7$.\cite{om00} This result showed that the gas layer flaring method
does not always result in the systematically highly-flattened halos
that the first two applications had, apparently by chance, returned.
Further, we were able to show that the result was consistent with the
constraint on halo flattening that one can derive from an analysis of
stellar kinematics perpendicular to the Galactic plane in the Solar
neighborhood: in that case, one uses a similar hydrostatic balance
between the stellar velocity dispersion and the pull of gravity toward
the plane to derive the total mass near the plane in the Solar
neighborhood; after subtracting the contribution from stars and gas in
this region, one obtains the amount of dark matter, and hence can
derive the flattening of the halo.

Returning for a moment to the shape of the halo in the planes of disk
galaxies (Section~\ref{sec:par}), it is worth noting that the
technique described above for determining the Milky Way's rotation
curve depends on the assumption of axisymmetry.  If this assumption
turns out to be invalid, one makes a different error in the derived
rotation curve than one would make using more conventional standard
candle methods for estimating the rotation curve.  Interestingly, there
is a small difference between the rotation curves derived by these two
methods, which has been used\cite{kt94} to estimate that the potential
in the plane of the Milky Way has an ellipticity of $\epsilon \sim
0.07$.

\begin{figure}[t]
\hfill \psfig{figure=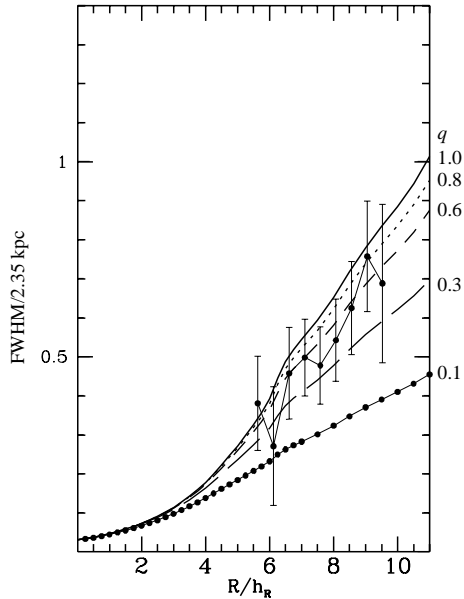,height=3.3in} \hfill{} 
\caption{The thickness of the HI layer in NGC~3198.  Observed
values\protect\cite{s97} have been fitted by dark halo models with a variety
of longest-to-shortest axis ratios, $q$.\label{fig:ngc3198}}
\end{figure}

Further confirmation that the gas layer flaring method does not return
systematically highly flattened halos has come from recent work that
we have undertaken on NGC~3198.  As the preliminary results presented
in Figure~\ref{fig:ngc3198}\ show, the variation in the thickness of the
HI layer with radius in this galaxy is not consistent with a highly
flattened halo, but favors a shortest-to-longest axis ratio of $\sim
0.6$.  

\begin{figure}[t]
\hfill \psfig{figure=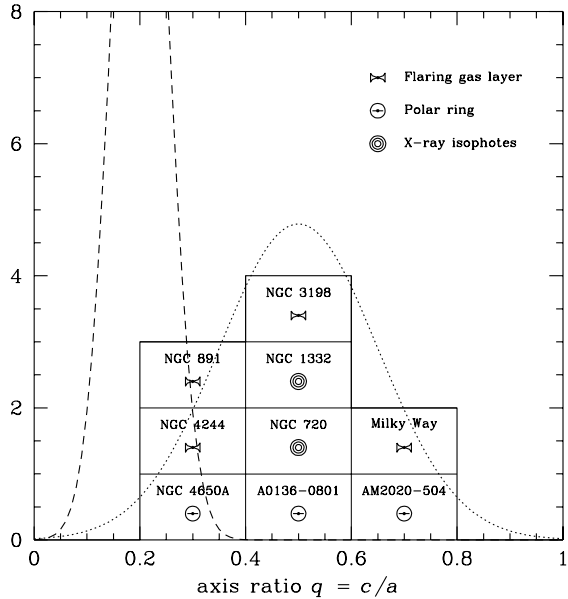,height=3.3in} \hfill{} 
\caption{Distribution of measured shortest-to-longest axis ratios of
galaxies' dark halos using the available credible techniques.  The
name and method for each galaxy is indicated.  The dotted line shows
the distribution predicted by cold dark matter simulations, while the
dashed line shows the distribution expected if cold molecular gas were
to make up the dark matter.\label{fig:shapehist}}
\end{figure}

\section{Summary}
Studying different aspects of the kinematics of HI gas in disk
galaxies provides us with a range of tools for probing the overall
distribution of mass in these systems.  Since the gas extends to large
radii where the mass is dominated by dark matter, these tools are well
suited to studying the shapes of galaxies' dark halos.  In the planes
of disk galaxies, the distribution seems to be very close to
axisymmetric, with some indications of an ellipticity in the potential
of $\sim 0.05$.  Perpendicular to the plane, a broader range of
flattenings is apparent: as Figure~\ref{fig:shapehist}\ summarizes,
shortest-to-longest axis ratios ranging from $\sim 0.2$ to $\sim 0.6$
are all found.  Further, the results from HI seem to be consistent
both with those obtained from studies of polar ring galaxies (so the
fears expressed in Section~\ref{sec:perp} seem to be unfounded) and
those derived from the X-ray isophotes of elliptical
galaxies.\cite{b01}

Using the data in Figure~\ref{fig:shapehist}, one can begin to make
quantitative comparisons with theory.  As the figure shows, the
observations are quite compatible with the predicted halo flattenings
that should occur when galaxies form in standard cold dark matter
cosmologies,\cite{d94} but are inconsistent with the highly-flattened
mass distribution that one would predict if all the dark matter were
made up from cold molecular material.\cite{petal94}  However, perhaps
the most notable fact apparent from Figure~\ref{fig:shapehist} is
quite how few reliable measurements of halo shape have actually been
made.  The challenge for the next few years is to populate this
histogram with sufficient new data for it to become a powerful
diagnostic for testing theories of galaxy formation and evolution.

\section*{Acknowledgements}
It is a pleasure to acknowledge Rob Olling's driving contribution to
much of the original research described in this review.

\end{document}